# Genetic Algorithms for Searching a Matrix of Metagrammars for Synthesis

Yixuan Li
University of Edinburgh

Federico Mora
University of California, Berkeley

Elizabeth Polgreen
University of Edinburgh

Sanjit A. Seshia
University of California, Berkeley

**Abstract**

Syntax-guided synthesis is a paradigm in program synthesis in which the search space of candidate solutions is constrained by a syntactic template in the form of a grammar. These syntactic constraints serve two purposes: constraining the language to the space the user desires, but also rendering the search space tractable for the synthesizer. Given a well-written syntactic template, this is an extremely effective technique. However, this is highly dependent on the user providing such a template: a syntactic template that is too large results in a larger search space and slower synthesis, and a syntactic template that is too small may not contain the solution needed. In this work, we frame the space of syntactic templates as a *matrix* of rules, and demonstrate how this matrix can be searched effectively with little training data using simple search techniques such as *genetic algorithms*, giving improvements in both the number of benchmarks solved and solving time for the state-of-the-art synthesis solver.

## 1 Introduction

Good grammars are essential for the scalable performance of program synthesis tools, particularly for synthesis solvers based on enumerative search. Grammars that are too low-level and too large result in large search spaces, and grammars that are too small and too high-level may omit valid solutions. Writing these good grammars is a difficult, and yet essential skill for a user of program synthesis engines. This is particularly true when using tools designed to accept the syntax-guided synthesis input language, SyGuS-IF [11], which is a low-level language similar to that of SMT-lib [5].

In this work, we present a matrix-based representation of the search space of context-free grammars for syntax-guided synthesis that is convenient to search, and easy to infer additional information from. We call this representation a metagrammar matrix, because it is a matrix that is capable of generating a family of grammars. We demonstrate that applying simple search techniques such as genetic algorithms to this search space is straightforward and effective, resulting in a matrix that generates grammars that speed up the state-of-the-art enumerative synthesizer by up to 50%.

In future work, we plan to explore using these matrix representations to infer minimal grammars and logical relations between rules in the grammars, as well as applying these to domains like verified lifting [9] where a user-provided high-level grammar is essential for the success of the synthesiser.

***Related work.*** Genetic programming is a technique for generating programs directly using a genetic algorithm [1] though without logical specifications. Genetic algorithms have been applied directly to the program synthesis problems with specifications [7][12], but they typically underperform in comparison to enumerative methods.

Previous work applies deep learning to filter grammars for programming-by-example problems(PBE) [10]. This deep learning is possible only because large training sets can be generated for PBE problems. We rely on lightweight machine learning methods like genetic algorithms, which require less training data, which allows our approach to go beyond PBE and be applied to general logical specifications. We have previously applied gradient descent [6] to the same problem, but we found that the search was limited to simple grammars with a single non-terminal per type, required more training iterations and does not achieve the same results. There are many synthesis techniques that benefit from using a coarser-grained representation of the search space, instead of the low-level grammars used by syntax-guided synthesis, noticeably component-based synthesis [8]. Our metagrammar search is a way to potentially automatically learn these components from a low-level grammar.

## 2 Background

***Syntax-guided synthesis.*** A SyGuS [2] problem consists of two constraints, a semantic constraint given as a formula $\sigma$, and a syntactic constraint in the form of a context-free grammar $G$, often referred to as the syntactic template. The computational problem is then to find a function $f$, that is permitted by a grammar $G$, and such that $\forall \vec{x}\ \sigma(f, \vec{x})$ is valid, where $x$ is the set of all possible inputs to $f$. The grammar $G = (V, N, R, S)$ is a context-free grammar, where $V$ is a set of symbols in the theory, $N$ is a set of non-terminals, $R$ is a set of production rules such that $R : (N \cup V)^* \mapsto (N \cup V)^*$ and $S$ is the start-symbol.

$G$ must also ensure that every sentence generated by the grammar is well-sorted and well-formed for the logic in question. For example, a grammar for synthesizing linear integer arithmetic programs must guarantee that no two integer variables are multiplied together in any term of the program.

*Enumerative SyGuS solvers.* SyGuS problems are commonly solved by the family of algorithms described by CounterExample Guided Inductive Synthesis(CEGIS) [14]. CEGIS alternates between two phases: a synthesis phase looks for a candidate program that works for a subset of inputs $\vec{x} \in I$; then a verification phase checks the candidate over the full set of program inputs. The synthesis phase is typically implemented using enumerative techniques [2, 3, 13], which enumerate through programs in the language $L^G$ until they find one that works for the subset of inputs $\vec{x} \in I$. In general, the more programs that an enumerative solver must enumerate through, the longer it takes to solve a synthesis problem. Applying synthesis in the real world often requires expert knowledge to write custom heuristics to make enumerative search effective in a specific domain, which means combining the expertise of both the synthesis expert and the domain expert.

*Genetic Algorithms.* A genetic algorithm is a learning technique based on the idea of natural selection. The algorithm begins with an initial candidate population, selects the best members of the population using a fitness function, and then repeatedly applies "cross-over" and "mutation" functions to these members to generate the next generation of the population.

## 3 A Matrix of Metagrammars

We represent the space of possible grammars with a *metagrammar*, i.e., a grammar of grammars. This can be thought of as a set of rules for generating a grammar. We represent this search space as a matrix, denoted $M_g$, where rows correspond to non-terminals and columns correspond to production rules. The entry in row $i$, column $j$ dictates whether a rule is included in the resulting grammar. Specifically, an entry of 1 indicates the rule is included, 0 indicates it is not included, and ✗ indicates that a rule is not valid for that non-terminal and cannot be included. An instantiation of $M_g$ is a matrix of 1's, 0's and ✗'s and corresponds to a single context-free grammar, denoted by $G$.

$$G = \begin{matrix} & R^1_{T_1} & R^2_{T_1} & R^1_{T_2} & R^2_{T_2} \\ N^1_{T_1} & 1 & 1 & \times & \times \\ N^2_{T_1} & 1 & 1 & \times & \times \\ N^1_{T_2} & \times & \times & 1 & 0 \\ N^2_{T_2} & \times & \times & 1 & 1 \end{matrix},$$

where $R^j_{T_i}$ indicates the $j^{th}$ rule that generates type $T_i$ and $N^k_{T_i}$ indicates the $k^{th}$ non-terminal of type $T_i$, and $(R^j_{T_i}, N^k_{T_i})$ is a valid rule- non-terminal pair. Multiple non-terminals of the same type allow grammars to specify finer-grained limits on the depth or operations used in sub-expressions, as opposed to simple grammars with a single non-terminal per type. We construct the metagrammar such that the first row of the matrix corresponds to the start symbol $S$.

### 3.1 Searching the space of metagrammars

The representation of a metagrammar opens up the possibility of multiple different search techniques. A natural first fit is genetic algorithms. In this section, we describe the details of the genetic search we implement, shown in Algorithm 1. We start with a metagrammar matrix that comprises of 2 non-terminals with types that appear in the specification, and a production rule per operator in the grammar of the corresponding theory.

We also add production rules that include the arguments to the synthesis function, basic constants 0 and 1, any constants that appear in the specification, and any helper functions that appear in the SyGuS benchmark. We initialize the gene pool with 60 randomly generated instances of this metagrammar matrix.

In theory, we are not limited to 2 non-terminals per type, but this restriction does prevent us from randomly generating some spurious grammars, e.g., a grammar with a large number of non-terminals that are not used in any production rule. A possible preprocessing step before evaluating the fitness is to eliminate the semantically redundant rules in our grammar. However, this could entail longer programs and increased computational complexity in finding solutions.

#### 3.1.1 The fitness function.
We select parents of the next generation using a fitness function, which receives a member of the population and returns a number that tells the fitness of that population member. We select a fitness function that trades off both runtime and the number of benchmarks solved: $f = n * \sum_{i=1}^{N}(T - t_i)$, where $n$ is the number of benchmarks solved, $N$ is the total number of benchmarks, $T$ is a timeout and $t_i$ is the runtime for solving the $i^{th}$ benchmark.

#### 3.1.2 Cross over.
The scattered crossover function takes a set of parents and enumerates through pairs, generating offspring from them until a population limit is reached. If our random binary matrix is [1 0 1 1; 0 1 1 1], where 0 means the gene is taken from the first parent and 1 means the gene is taken from the second parent. Parents $G_{p1}$, $G_{p2}$ will generate offspring $G_O$:

$$G_{p1} = \begin{bmatrix} 1 & 0 & \times & \times \\ \times & \times & 1 & 0 \end{bmatrix}, G_{p2} = \begin{bmatrix} 0 & 1 & \times & \times \\ \times & \times & 0 & 1 \end{bmatrix};$$

$$G_O = \begin{bmatrix} 0 & 0 & \times & \times \\ \times & \times & 0 & 1 \end{bmatrix}.$$

#### 3.1.3 Mutate.
We apply the mutation function randomly to the children that result from the crossover function.

| Benchmarks Set | Total | Default Grammar | | | Metagrammar | | | Improvement | |
|---|---|---|---|---|---|---|---|---|---|
| | | # Solved | Avg. Time(s) | % Solved | # Solved | Avg. Time(s) | % Solved | Time(s) | % solved |
| Invertibility Conditions | 100 | 59 | 1.56s | 59.0% | 79 | 1.89s | 79.0% | -0.33s | 20.0% |
| Conditional Inverses | 100 | 87 | 2.04s | 87.0% | 86 | 0.11s | 86.0% | 1.93s | -1.0% |
| PBE | 690 | 248 | 7.39s | 36.0% | 670 | 2.71s | 97.1% | 4.68s | 61.1% |
| LIA | 50 | 5 | 4.27s | 10.0% | 5 | 1.83s | 10.0% | 2.44s | 0.0% |
| Total(BV) | 890 | 394 | 3.66s | 44.2% | 835 | 1.57s | 93.8% | 2.09s | 49.6% |
| Total(General) | 940 | 399 | 3.82s | 42.4% | 840 | 1.64s | 89.4% | 2.18s | 47.0% |

**Table 1.** The number of benchmarks solved and average solving time in seconds for default grammar and metagrammar-generated grammar, and the performance improvements achieved.

Given a matrix $G_1$, the mutation function randomly mutates $G_1$ to $G_2$ by randomly flipping entries in the matrix from 1 to 0 or vice versa. An example mutation is shown below.

$$G_1 = \begin{bmatrix} 1 & 0 & \times & \times \\ \times & \times & 1 & 0 \end{bmatrix} \Rightarrow G_2 = \begin{bmatrix} 1 & 1 & \times & \times \\ \times & \times & 1 & 0 \end{bmatrix}.$$

**Algorithm 1** Metagrammar Search

```
function SEARCH(f, M_g, benchmarks B)
    Pop ← genRandGrammars(M_g, 60)
    generation ← 0
    while generation ≤ MAX do
        Parents ← bestN(f, B, Pop, 15)
        Children ← CrossOver(Parents, 60)
        Pop ← Mutate(Children)
        generation ← generation + 1
    end while
    return BestN(f, B, Pop, 1)
end function
```

## 4 Evaluation

We evaluate our approach on benchmarks taken from the SyGuS competition. We use an initial population of 60, a set of 15 parents and run the algorithm for 100 generations. We split the benchmark set randomly into testing and training data, with 60 benchmarks selected for training. We envision this kind of search would be used when a user has a specific application in mind, and a corpus of benchmarks from that application, where they can train on a small subset and then use the resulting metagrammar on the remainder of the set. Consequently, we split the benchmarks into four broad categories shown in Table 1.

We run the genetic algorithm to generate a metagrammar for each category, and then use CVC5 [4] version 1.0.2 to solve the test set benchmarks using the grammars generated by that metagrammar. For comparison, we also run CVC5 on a set of benchmarks with the full grammar given if every non-✗ entry in the metagrammar is set to 1.

### 4.1 Summary and Discussion

Overall CVC5 solved 449 benchmarks with metagrammar-generated grammars that it could not solve with the full grammar. It failed to solve 8 benchmarks with metagrammar generated grammars that it could solve with the full grammar (3 in bitvector benchmarks and another 5 in LIA benchmarks). We solve 2.1× more benchmarks with the metagrammar, and the average solving time improved by 47%. Previous work [6] achieved a performance improvement of 1.4× more benchmarks.

Performance is not uniform across the categories: we perform well on bitvectors but improvement on Linear Integer Arithmetic is minimal. We hypothesize this is due to the smaller number of operators in LIA. We also fail to improve the number of benchmarks solved in the conditional inverse category, although we note that the baseline with the default grammar already performs particularly well in this category.

The metagrammar may sacrifice the ability to solve one or two benchmarks that the default grammar can solve, in order to enable solving of a much larger new set (this is particularly noticeable in the invertibility conditions where there are 2 benchmarks unsolved with the metagrammar and solved with the default grammar, but 22 new benchmarks solved by the metagrammar). This is due to the fitness function we have chosen, and the behaviour could be altered if a different fitness function were provided.

## 5 Conclusions and Future Work

We have demonstrated that the matrix-based representation of a metagrammar allows easy application of simple effective search techniques to the space of grammars. We believe this is essential to lifting formal synthesis techniques to applications that require more scalable synthesis, such as program lifting [9], and which are currently only achievable when the user provides the synthesizer with custom-written high-level Domain Specific Languages. In future work, we hope to explore other simple search techniques and to expand the expressiveness of these matrix metagrammars, including automatically inferring logical relations between matrix elements. We also believe that this simple way of writing a metagrammar would allow users to easily provide more elaborate grammars for synthesis problems.